\documentclass[a4paper,10pt]{article}

\usepackage[margin=25mm]{geometry}
\usepackage{times}
\usepackage{graphicx}
\usepackage{caption}
\usepackage{hyperref}
\usepackage{amsmath}
\usepackage{xcolor}
\usepackage{fancyhdr}
\usepackage{titlesec}
\titleformat{\section}{\bfseries\large\uppercase}{\thesection.}{1em}{}
\titleformat{\subsection}{\bfseries\normalsize}{\thesubsection}{1em}{}
\titleformat{\subsubsection}{\itshape\normalsize}{\thesubsubsection}{1em}{}

\renewenvironment{abstract}{
  \begin{center}
    {\large\bfseries Abstract \par} 
  \end{center}
}{\par\vspace{2em}}

\usepackage[numbers, compress]{natbib}

\newcommand{\lens}{\vec\theta_{L}}
\newcommand{\light}{\vec\theta_{l}}
\newcommand{\bbh}{\vec\phi_{BBH}}
\newcommand{\bbl}{\vec\phi_{L}}

\newcommand{\fixed}[1]{\textcolor{black}{#1}}

\newcommand{\mnras}{\textit{MNRAS}}
\newcommand{\aap}{\textit{Astronomy \& Astrophysics}}
\newcommand{\pr}[1]{\textit{Phys. Rev. #1}}
\newcommand{\apjl}{\textit{ApJL}}
\newcommand{\apj}{\textit{ApJ}}
\newcommand{\jcap}{\textit{JCAP}}

\begin{document}

\title{\bfseries Joint Bayesian Source and Lens Reconstruction for Multi-messenger Binary Black Holes}
\author{Laura E. Uronen$^{1,2}$, Tian Li$^{3}$, Justin Janquart$^{4,5}$, Hemanta Phurailatpam$^{1}$, Jason S.C. Poon$^{1}$, \\ Thomas E. Collett$^{3}$, Léon V.E. Koopmans$^{2}$, Otto A. Hannuksela$^{1}$  \\
{\small $^{1}$Department of Physics, The Chinese University of
Hong Kong, Shatin, NT, Hong Kong, {\it laura.uronen@gmail.com}}\\
{\small $^{2}$Kapteyn Astronomical Institute, University of
Groningen, P.O Box 800, 9700 AV Groningen, The
Netherlands}\\
{\small $^{3}$Institute of Cosmology and Gravitation, University of
Portsmouth, Burnaby Rd, Portsmouth, PO1 3FX, UK}\\
{\small $^{4}$Centre for Cosmology, Particle Physics and
Phenomenology, Université Catholique de
Louvain,}\\
{\small Louvain-La-Neuve, B-1348, Belgium}\\
{\small $^{5}$Royal Observatory of Belgium, Avenue Circulaire, 3,
1180 Uccle, Belgium}}
\date{}
\maketitle
\thispagestyle{fancy}

\begin{abstract}
	If a gravitational wave event is lensed by a cluster or galaxy in our line-of-sight, it is expected that its host galaxy would also be lensed. 
	Therefore, connecting lensed gravitational wave events even without direct optical counterpart could be feasible by identifying matching lenses in electromagnetic data and surveys. 
	Seminal work has demonstrated the potential of this approach in LVK, Euclid, HST, JWST, and CSST mock data, motivating the need for a dedicated software package to perform such analyses in practice.
	Here, we present the alpha-version of \texttt{silmarel}, the first software package designed to bridge these cosmic signals and enable us analysis of real LVK gravitational-wave binaries together with telescope observations from instruments like \textit{Euclid} or \textit{Hubble} Space Telescope, and the future of multimessenger binary black hole lensing.
\end{abstract}

\section{Introduction}

The theory of general relativity (GR) posited a Universe gave rise to two new branches of observational astrophysics: gravitational lensing, the bending of light by massive structures, and gravitational waves (GWs), the ripples of spacetime itself. 
For decades, lensing has been a cornerstone of electromagnetic (EM) observations, while since their first detection in 2015 \cite{gw150914}, GWs are now heard by the LIGO-Virgo-KAGRA observatories \cite{lvk,aligo,o4ligo,o4aligo,avirgo,kagra,kagrab,kagrac} at an ever-accelerating rate \cite{gwtc4}.
Now, a new frontier lies at the union of these two pillars of GR: strong-lensing of GWs \cite{schneider1992}, with the first detection having been forecasted by several independent groups to take place as further detector upgrades are implemented~\cite{ng2017, li2018, oguri2018}. 
This is amongst the reasons there has been a global effort to search for lensed GWs in the LVK data and
implement strategies to look for it \cite{hannuksela2019, li2023, dai2020, liu2021, mcisaac2020, o3alensing, o3lensing, o3followup, o4alensing}. 

In anticipation of the first detection of lensed GWs, it is now crucial that our methods be ready to meet the challenge. 
If detected, GW lensing has applications in 
probing cosmology \cite{liao2017, hannuksela2020, finke2021, jana2023, balaudo2023, narola2024, wempe2024a, jana2024, maity2024, chen2026} and 
the large-scale structure of the Universe \cite{mukherjee2020, savastano2023, vujeva2025} 
and the nature of dark matter \cite{jung2019, urrutia2021, basak2022, tambalo2023, jana2024b, barsode2024} 
among others \cite[and references in both]{ o4alensing,smith2025}.
One particular approach breaches across fields: multi-messenger lensing using dark binaries \cite{hannuksela2020, wempe2024b, uronen2024, chen2022, shan2025}. 
Until recently, traditional multi-messenger lensing studies relied on luminous mergers like GW170817 \cite{gw170817a}, where the connection between light and GW is intrinsic \cite{gw170817b}. 
If, however, we could bring the entire population of lensed events—including binary black holes (BBHs)—into the equation, our understanding of the Universe can only be deepened and enriched. 
Since BBH formation theories, whether formed directly from stellar binaries or through dynamic mergers in clusters or AGNs, typically result in BBHs formed inside bright host galaxies, we aim to connect the lensed GW merger to the light of the galaxy they reside in. 

Here, we introduce \texttt{silmarel}\footnote{github.com/laurauronen/silmarel} (Statistical Inference for Lensing Multi-messenger Assocation, REconstruction, and Localisation), a Bayesian framework designed to unite lensed GW events and the telescope observations of lensed galaxies in the sky to enable joint multi-messenger analysis of these unique events. 

We describe the mathematical and physical foundation of this method, show a simple demonstration with simulated data, and chart the path for future expansion as we prepare to see the cosmos in a new light.

\section{Methods}

\subsection{Electromagnetic lens reconstruction}

A single equation defines, for the most part, all of gravitational lensing: the lens equation. 
It relates the original position of our source, $\vec\beta$, 
to the final position of the image of the source, $\vec\theta$, 
by means of the angle the beam of light has been deflected by the lens, $\vec\alpha$ \cite{schneider1992}:
\begin{equation}
    \vec\beta = \vec\theta - \alpha(\vec\theta) \,. 
\end{equation}
For strong-lensed systems such as galaxy or cluster-scale lenses with sources like background galaxies, quasars, or supernovae, this will usually result in multiple ‘copies’ (images, \textit{j}) of the source, \fixed{and where extended source will be differentially distorted, each point source image will have} a magnification $\mu_j$, time delay $\Delta t_j$ (for time-variable sources), and image parity, which results in a ‘flipping’ of the image depending where on the image plane it is \cite{wambsganss1998}. 

Since the light has been deflected by structure between source and observer, the first step is reconstructing the lens system.
Such lens reconstruction is usually done through forward-modelling, where a morphology is initially proposed for the lens and source, and are then iterated and adjusted to reduce residuals and match what our image has shown. 
A number of tools (e.g. \texttt{lenstronomy} \cite{lenstronomy}, \texttt{herculens} \cite{herculens}, \texttt{PyAutoLens} \cite{autolens}) explore this process using different of source/lens profiles and morphologies or inference/computational methods. 

\subsection{Strong lensing gravitational-wave parameter estimation}

Corresponding effects are observable for any lensed GW. 
Taking an unlensed GW waveform $\tilde{h}_U(f)$ as a function of frequency $f$, 
the resulting lensed waveform $\tilde{h}_{L,j}(f)$ for a given image \textit{j} is \cite{takahashi2003}: 
\begin{equation}
    \tilde{h}_{L,j}(f) = \sqrt{\mu_j}e^{-2\pi i(f\Delta t_j - \frac{n_j}{2})} \tilde{h}_{U}(f),
\end{equation}
where the other terms are defined above and the phase term $n_j$, called the ‘Morse factor’, corresponds to the image parity for GWs.
All other parameters must match between images, such as the observed chirp mass or spins of the GW event, since strong lensing is monochromatic. 

However, when a lensed GW is first observed by the LVK, we do not know its real luminosity distance, nor the absolute time of arrival, because these effects are degenerate with the lensing magnification and time delay, respectively~\cite{dai2017, haris2018}.
So for a single image, we cannot measure absolute time delays or magnifications inflicted by lensing—we can only measure the relative ones between GW ‘images’ (see however~\cite{poon2024, seo2024} which demonstrate that the magnifications and time delays could be determined for quadruply lensed events when a simple galaxy lens model is defined prior to the analysis; here we are at first agnostic to the lens model). 
Such techniques are established using joint parameter estimation (JPE)~\cite{dai2020, liu2021, haris2018, janquart2021, janquart2023, lo2023, barsode2025}: We match GW events to one another by probing whether the BBH parameters match \fixed{and sky localisations overlap}, with some relative lensing parameters between the images,  
This can be done for example by packages like \texttt{GOLUM} \cite{janquart2021, janquart2023} and \texttt{hanabi} \cite{lo2023}.
If these events match, the GWs may be lensed copies of one another, though rigorous statistical analysis is required to confirm this~\cite{Hannuksela2025,Barsode2025}. 

\subsection{Joint lens reconstruction}

From Sec. 2.1 and 2.2, it is clear that for any GW-emitting source with a corresponding and observable EM-source, the lens properties must directly link the two datasets together. 
However, due to the limited number of constraints from the data coming from the lensing parameters $\{ \mu_j, \Delta t_j, n_j\}$ measured relative between GW images, the high-dimensional nature of lens reconstruction means that gravitational-wave-only reconstructions suffer from the mass-sheet and similarity transformation degeneracies when trying to reconstruct lens parameters on the GW side alone \cite{poon2024}. 
The only suitable solution then would be to identify the lens from those observed in the sky.

A bright binary neutron star, for example, may be observable directly and its host galaxy and the acting lens pin-pointed through this observation\cite{nicholl2025}.
On the contrary, black holes and their collisions are not expected to emit observable light, so their corresponding signatures risk being non-existent. 
However, if these mergers reside inside galaxies, it is reasonable to expect that the same lens will equivalently act upon both the host and BBH. 

While we can use methods like GW-only lens reconstruction to reduce the list of lenses in the sky patch, it is not precise enough to identify the unique lens. 
However, the two datasets are complementary. 
Galaxy lenses do not provide time-delay information and GWs provide it in extremely high resolution, while where GWs lack spatial resolution, EM telescopes can provide this. 
GWs also act as standard sirens\footnote{GWs provide the luminosity distance, which for lensing can also be used to directly measure the absolute magnification if the source redshift is retrieved from the electromagnetic band.}, are immune to dust extinction and microlensing~\cite{cheung2021,yeung2023}, and have different selection effects than EM observations, which can help break degeneracies in the lens reconstruction and provide complementary information about the lens and source~\cite{poon2024,cheung2021,yeung2023}. 
Therefore, the two datasets can be combined to identify the common lens through a joint lens reconstruction.

For some given lens, we can compute the Bayes' factor of whether the EM lens data $d_{EM}$ is associated ($\mathcal{H}_A$) with the GW data $d_{GW}$ or not ($\mathcal{H}_N$) as shown in Ref.~\cite{hannuksela2020, wempe2024b, uronen2024}. 
This provides the joint likelihood: 
\begin{equation}
    p(d_{GW}, d_{EM}, d_{spect}| \mathcal{H}_A) = \int p(d_{GW} | \bbh, \bbl(\lens)) p(d_{EM} | \lens, \light) p(d_{spect} | z_l) p(\vec\theta_{all})d\vec\theta_{all}\,,
\end{equation}
where $\bbh$ are the unlensed BBH parameters, 
$\bbl$ are the lensed BBH parameters which arise as a result of the lens mass $\lens$, 
$\light$ corresponds to a combination of source and lens light parameters, giving the total set of parameters $\vec\theta_{all}$. 
We also include spectroscopic data \fixed{$d_{spect}$} of the lens redshift, which can typically be measured to high precision. 
This likelihood is the full, joint multi-messenger likelihood we use to combine EM and GW data. 

However, GW JPE is already costly without EM data. 
There are 15 unlensed $\bbh$ parameters, 
a single lensed GW image analysis counts $N_{\bbh}$+1 parameters, and each additional GW image adds an extra 3 parameters. 
The waveform generation is computationally intensive, and depending on the waveform model and the priors JPE can take anywhere from $\mathcal{O}$(hours — weeks). 
With high-dimensional EM lens reconstruction and the current resources, the likelihood becomes computationally difficult. 
We must therefore find an alternative solution.

\section{Silmarel}

We may reformulate the joint GW–EM lensing problem by replacing the computationally prohibitive likelihood evaluations with posterior-induced effective likelihoods to avoid the need to generate full waveforms: 

\begin{enumerate}
    \itemsep0em 
    \item Many of the parameters relate to mass or spin, which are unaffected by lensing.
    \item Only the observed or effective luminosity distance and arrival time of each image have been affected by lensing, through relative magnification $\mu_{rel}$ and relative time delay between images $\Delta t$ respectively.
    \item The image parity information $n_j$ provides little to no contribution to the lens reconstruction. 
    \item For every GW pair, JPE analyses are already routinely completed as part of LVK catalogue releases \cite{hannuksela2019,o3alensing,o3lensing,o3followup,o4alensing}: therefore, posteriors for every $\bbh, \bbl$, noted $p(\bbh,\bbl | d^{all}_{GW})$, already exist.
    \item The EM likelihood can be approximated as unaffected by GW information.
\end{enumerate}

Using these, we can reduce the problem. By recasting the GW data reduced to the set of posteriors affected by lensing $\{\vec\mu_{rel}, \Delta \vec t\}$ as functions of the lens parameters, and including the EM/spectroscopic likelihood independent of GW information \cite{wempe2024b, uronen2024}: 

\begin{equation}
    p(d_{GW}, d_{EM}, d_{spect}| \mathcal{H}_A) = \int p(\vec d_{GW} | \vec\mu_{rel}, \Delta \vec t) p(d_{EM}, d_{spect} | \lens, \light, z_l, z_s) p(\vec\theta'')d\vec\theta'',
\end{equation}

\begin{figure}
    \centering
    \includegraphics[width=0.9\linewidth]{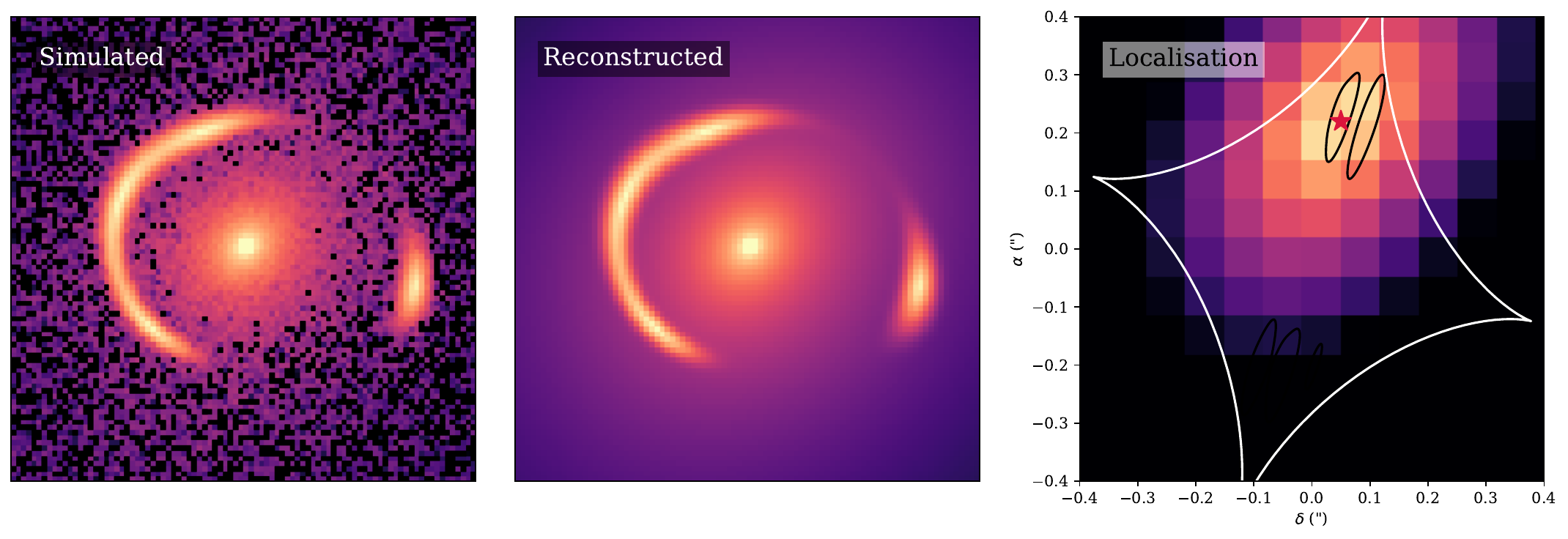}
    \caption{Reconstruction using the \texttt{silmarel} likelihood with \texttt{lenstronomy} lens reconstruction. From left to right, the plots show the initial simulated observation, the final reconstructed lens, and the localisation in the source plane showing the posteriors (black contours), true source (pink star), the source profile and lens caustics.}
    \label{fig:method}
\end{figure}

with $\vec\theta''$ the final reduced set of parameters, \fixed{$\{\lens, \light, \mu_{rel}, \Delta t, z_l, z_s\}$}. As we can obtain our model values for $\mu_{rel}, \Delta t$ for each image from the lens parameters and redshifts, and we already have posteriors for these parameters, the GW likelihood can therefore become: 

\begin{equation}
    \log p(d_{GW} | \fixed{\vec\mu_{rel}, \Delta \vec t}) = - 0.5 \sum_j \exp \left[ \frac{(\overline{\mu_{rel,j}} - \mu^{model}_{rel})^2}{\sigma_{j,\mu}^2} + \frac{(\overline{\Delta t_j} - \Delta t^{model})^2}{\sigma_{model, t}^2} \right],
    \label{eq:silm-ll}
\end{equation}

replacing expensive likelihood evaluations with posterior-based Gaussian likelihood terms around the median time delay $\overline{\Delta t}$\footnote{\fixed{Formally, the time delay the likelihood is $p(d_{GW} | \Delta\vec t_{EM}) = p(d_{GW} | \Delta\vec t_{GW})p(\Delta\vec t_{GW} | \Delta t_{EM})$, with GW posterior $
\Delta\vec t_{GW}$. The first term becomes a delta function around the median, $\overline{t_{j}}$, while the second term is the Gaussian likelihood with EM modelling uncertainty.}} and magnification $\overline{\mu}$.\footnote{\fixed{The uncorrelated Gaussian for magnification is a first-order approximation and does not include potential covariances between GW posteriors such as luminosity distance, inclination and magnification ratios, which will be added in future work.}}

Although the uncertainty for magnification is set by the uncertainty in the GW data, we cannot use the uncertainty in the GW data for time delays. The LVK measures arrival times to a precision of $\mathcal{O}$(ms), and the resulting data uncertainty is also of this order for time delays. \fixed{However, EM imaging and common lens mass models cannot resolve or reconstruct the truth deflection lens potentials to the precision GWs measure.} Therefore, time delays by lensing mass that GWs are sensitive to and can reproduce are often not observable in the EM nor reconstructed in EM lens models. Therefore, using the true GW uncertainty could lead to inability to identify the correct lens as none of them can reproduce the time delays to the precision GWs can. We must therefore assign a larger uncertainty to the GW time delay, to prevent missing associated lenses caused by precision EM imaging cannot meet. 

Eq.~\ref{eq:silm-ll} is the likelihood used in \texttt{silmarel}. The benefit of this approach is that it provides us also with a direct measure of the localisation of the GW, since the source position is sampled. The \texttt{LenstronomyLikelihood} class implements \texttt{lenstronomy} lens reconstruction followed by the GW localisation, which allows independence between the EM and GW reconstruction stages and can speed up analysis—for example, for pre-existing lens reconstructions, one can instead directly run the GW-only localisation without need to re-do the lens reconstruction. This provides a “fast likelihood” for localisation. We show a demonstration of this in Fig.~\ref{fig:method}.

\section{Future work and conclusions}

We have presented \texttt{silmarel}, a package for joint GW and EM multi-messenger lensing data analysis. Through GW posteriors existing as direct LVK data products and EM lens reconstruction tools, we can combine the two datasets to obtain an association between host and GW, as well as localise the GW inside of its host galaxy to a precision several orders of magnitude greater than typical LVK sky localisations. Currently, \texttt{silmarel} consists of a fast estimator that cuts the likelihood in stages to complete the EM reconstruction and GW localisation separately.

Where currently the fast estimator has been implemented, \texttt{silmarel} aims in the future to allow wider agnosticity in relation to choice of lens reconstruction with work to expand the EM reconstruction to combine seamlessly with other reconstruction frameworks such as \texttt{herculens, PyAutoLens,} which will also allow for an integrated likelihood that combines the GW and EM reconstructions directly, rather than in separated stages.

However, where \texttt{silmarel} is the first step towards joint multi-messenger, the complete inference remains an exciting challenge mathematically, physically, and computationally. To this end, exploring avenues like approaches founded in machine-learning to speed up GW waveform generation and data analysis, and the use of inference with fairly high dimensionality, are necessary future directions for the field.

\end{document}